\documentclass[a4paper,11pt]{article}
\usepackage[left=20mm,right=20mm,top=20mm,bottom=20mm]{geometry}
\usepackage{amsmath}
\usepackage{amsfonts}
\usepackage{amssymb}
\usepackage{amsbsy}
\usepackage{color,graphics}
\usepackage{setspace}
\usepackage{epsfig}
\usepackage{cite}
\usepackage{graphicx}
\usepackage{latexsym,multicol}

\graphicspath{{./SEIN-figs/}}
\begin{document}
\title{Likely equilibria of stochastic hyperelastic spherical shells and tubes}
\author{L. Angela Mihai\footnote{School of Mathematics, Cardiff University, Senghennydd Road, Cardiff, CF24 4AG, UK, Email: \texttt{MihaiLA@cardiff.ac.uk}}
	\qquad Danielle Fitt\footnote{School of Mathematics, Cardiff University, Senghennydd Road, Cardiff, CF24 4AG, UK, Email: \texttt{FittD@cardiff.ac.uk}}
	\qquad Thomas E. Woolley\footnote{School of Mathematics, Cardiff University, Senghennydd Road, Cardiff, CF24 4AG, UK, Email: \texttt{WoolleyT1@cardiff.ac.uk}}
	\qquad Alain Goriely\footnote{Mathematical Institute, University of Oxford, Woodstock Road, Oxford, OX2 6GG, UK, Email: \texttt{goriely@maths.ox.ac.uk}}
}	\date{October 13, 2018}
\maketitle

\begin{abstract}
In large deformations, internally pressurised elastic spherical shells and tubes may undergo a limit-point, or inflation, instability manifested by a rapid transition in which their radii suddenly increase. The possible existence of such an instability depends on the material constitutive model. Here, we revisit this problem in the context of stochastic incompressible hyperelastic materials, and ask the question: what is the probability distribution of stable radially symmetric inflation, such that the internal pressure always increases as the radial stretch increases? For the classic elastic problem, involving isotropic incompressible materials, there is a critical parameter value that strictly separates the cases where inflation instability can occur or not. By contrast, for the stochastic problem, we show that the inherent variability of the probabilistic parameters implies that there is always competition between the two cases. To illustrate this, we draw on published experimental data for rubber, and derive the probability distribution of the corresponding random shear modulus to predict the inflation responses for a spherical shell and a cylindrical tube made of a material characterised by this parameter. \\
	
	
\end{abstract}

\section{Introduction}

The idealised model of an internally pressurised  hollow cylinder or sphere is instructive as it applies to many structures from living cells to blood vessels to aircraft fuselages \cite{goriely17,Vogel:1998}. For these structures to be serviceable they must be able to withstand and function at a certain level of internal pressure without damage. The finite symmetric inflation and stretching of a cylindrical tube of homogeneous isotropic incompressible hyperelastic material was initially studied by Rivlin (1949) \cite{Rivlin:1949:VI}. For an elastic spherical shell, the finite radially symmetric inflation was first investigated by Green and Shield (1950) \cite{Green:1950:GS} (see also \cite{Adkins:1952:AR,Shield:1972}). A general theory of possible qualitative behaviours for both the elastic tubes and the spherical shells was developed by Carroll (1987) \cite{Carroll:1987}, who showed that, depending on the particular material and initial geometry, the internal pressure may increase monotonically, or it may increase and then decrease, or it may increase, decrease, and then increase again. This formed the basis for further studies where these deformations were examined for different material constitutive laws \cite{Muller:2002:MS,Goriely:2006:GDBA,Zamani:2017:ZP}, and opened the way to the modelling of more complex phenomena \cite{Haas:2015:HG,Mihai:2015:MAG,Weickenmeier:2017:etal}. Recently, localised bulging instabilities in an inflated isotropic hyperelastic tube of arbitrary thickness, which were also observed in some materials, provided that the tube is sufficiently long \cite{Gongalves:2008:GPL}, were modelled and analysed within the framework of nonlinear elasticity in \cite{Fu:2016:FLF}.

Clearly, the behaviour of a structure depends on the inextricable relation between its material properties and its geometry. It is therefore of the utmost importance to use suitable constitutive models as confirmed by experience and  experiments. Unavoidably, uncertainties are attached to material properties. For natural and industrial elastic materials, uncertainties in the mechanical responses generally arise from the inherent micro-structural inhomogeneity, sample-to-sample intrinsic variability, and lack of data, which are sparse, inferred from indirect measurements, and contaminated by noise \cite{Bayes:1763,Farmer:2017,Hughes:2010:HH,Oden:2018}. Stochastic approaches are thus growing in importance as a tool in many disciplines, such as materials science, engineering, and biomechanics, where understanding the variability in the mechanical behaviour of materials is critical. For example, constitutive equations for soft tissues, including those based on statistical modelling for the evolution of the collagen network, are reviewed in \cite{Chagnon:2014:CRF}. and  statistical approaches applied to the mechanical analysis of rubberlike networks is presented in \cite{Kloczkowski:2002}. Further, methods for stochastic stability  motivated by applications to stochastic network  can be found in \cite{Foss:2004:FK}.

Within the nonlinear elasticity field theory, which is based on average data values and covers the simplest case where internal forces only depend on the current deformation of the material and not on its history, hyperelastic materials are the class of material models described by a strain-energy function with respect to the reference configuration \cite{goriely17,Ogden:1997,TruesdellNoll:2004}. For these materials, boundary value problems can be cast as variational problems, which provide powerful methods for obtaining approximate solutions, and can also be used to generate finite element methods for computer simulations. The tradition so far has been to only consider average values to fit deterministic models. More recently, the use of the information about uncertainties and the variability in the acquired data in nonlinear elasticity has been proposed by the introduction of \textit{stochastic hyperelastic models}. Specifically, stochastic representations of isotropic incompressible hyperelastic materials characterised by a stochastic strain-energy function for which the model parameters are random variables following standard probability laws were constructed in \cite{Staber:2015:SG}, while compressible versions of these models were presented in \cite{Staber:2016:SG}. Ogden-type stochastic strain-energy functions were then calibrated to experimental data for rubber and soft tissues in \cite{Mihai:2018a:MWG,Staber:2017:SG}, and anisotropic stochastic models with the model parameters as spatially-dependent random fields were calibrated to vascular tissue data in \cite{Staber:2018:SG}. These models  are based on the notion of  entropy (or uncertainty) defined by Shannon (1948) \cite{Shannon:1948,Soni:2017:SG}, and employ the maximum entropy principle for a discrete probability distribution introduced by Jaynes (1957) \cite{Jaynes:1957a,Jaynes:1957b,Jaynes:2003}. However, by contrast to the tremendous development and variety of stochastic finite element methods, which have been proposed and implemented extensively in recent years \cite{ArreguiMena:2016:AMMM,Babuska:2005:BTZ,Hauseux:2017:HHB,Hauseux:2018:HHCB}, at the fundamental level, there is very little understanding of the uncertainties in the mechanical properties of elastic materials under large strains. For these models, the natural question arises: \emph{what is the influence of the stochastic model parameters on the predicted elastic responses?} 

In recognition of the fact that a crucial part of assessing the elasticity of materials is to quantify the uncertainties in their mechanical responses under large deformations, in the present study, we determine the probability distribution of stable deformation for spherical shells and cylindrical tubes of stochastic isotropic hyperelastic material under radially symmetric inflation. For the deterministic elastic problem, involving isotropic incompressible materials, there is a critical parameter value that strictly separates the cases where inflation instability can occur, or not. However, for the stochastic problem, we find that, due to the probabilistic nature of the material law, there is always competition between the two cases. Therefore, we can no longer talk about `equilibria' but `likely equilibria' obtained under a given internal pressure with a given probability. Our stochastic elastic setting provides a general mathematical framework applicable to a class of stochastic hyperelastic materials for which similar results can be obtained. As a specific example, we refer to the experimental data for vulcanised rubber of Rivlin and Saunders (1951) \cite{Rivlin:1951:VII:RS}, from which we derive the probability distribution of the random shear modulus, and predict the inflation responses for a spherical shell and a cylindrical tube made of a material characterised by this parameter. We begin, in Section~\ref{sec:models}, with a detailed description of the stochastic elasticity framework. Then, in Sections~\ref{sec:shell} and \ref{sec:tube}, for stochastic spherical and tubes, respectively, first, we review the solution to the elastic problem under radially symmetric inflation, then we recast the problem in the stochastic setting and find the probabilistic solution. Concluding remarks and a further outlook are provided in Section~\ref{sec:conclude}.

\section{Stochastic isotropic hyperelastic materials}\label{sec:models}

In this Section, we summarise the stochastic elasticity framework developed in \cite{Mihai:2018a:MWG} using the  methodology and previous results of \cite{Staber:2015:SG,Staber:2016:SG,Staber:2017:SG}. This theoretical framework was used to analyse the role of stochasticity in stability problems in \cite{Mihai:2018b:MWG,Mihai:2018:MDWG}.

\subsection{Stochastic setting}

We briefly recall that a homogeneous hyperelastic material is characterised by a strain-energy function $W(\mathbf{F})$ that depends on the deformation gradient tensor, $\mathbf{F}$, with respect to the reference configuration \cite{goriely17,Ogden:1997,TruesdellNoll:2004}. It is typically characterised by a set of \textit{deterministic} model parameters, which then contribute to defining the constant elastic moduli under small strains, or the nonlinear elastic moduli, which are functions of the deformation, under large strains \cite{Mihai:2017:MG}. By contrast, a stochastic homogeneous hyperelastic material is described by a stochastic strain-energy function for which the parameters are \emph{random variables} that satisfy standard probability laws \cite{Mihai:2018a:MWG,Staber:2015:SG,Staber:2016:SG,Staber:2017:SG}. Specifically, each parameter is described in terms of its \emph{mean value} and its \emph{variance}, which includes information about the range of values about the mean value. The mean value and the variance are the most commonly used to provide information about a random quantity in many practical applications \cite{Hughes:2010:HH}. Here, we combine information theory with finite elasticity, and rely on the following assumptions \cite{Mihai:2018a:MWG,Mihai:2018b:MWG}:
\begin{itemize}
\item[(A1)] Material objectivity: the principle of material objectivity (frame indifference) states that constitutive equations must be invariant under changes of frame of reference. It requires that the scalar strain-energy function is unaffected by a superimposed rigid-body transformation (which involves a change of position) after deformation, i.e., $W(\textbf{R}^{T}\textbf{F})=W(\textbf{F})$, where $\textbf{R}\in SO(3)$ is a proper orthogonal tensor (rotation). Material objectivity is guaranteed by considering strain-energy functions defined in terms of invariants.

\item[(A2)] Material isotropy: the principle of isotropy requires that the scalar strain-energy function is unaffected by a superimposed rigid-body transformation prior to deformation, i.e., $W(\textbf{F}\textbf{Q})=W(\textbf{F})$ where $\textbf{Q}\in SO(3)$. For isotropic materials, the  strain-energy  function is a symmetric function of the  principal stretches of $\textbf{F}$, $\lambda_{1}$, $\lambda_{2}$, $\lambda_{3}$, i.e.,  $W(\textbf{F})=\mathcal{W}(\lambda_{1},\lambda_{2},\lambda_{3})$.

\item[(A3)] Baker-Ericksen inequalities: in addition to the fundamental principles of objectivity and material symmetry, in order for the behaviour of a hyperelastic material to be physically realistic, there are some universally accepted constraints on the constitutive equations. Specifically, for a hyperelastic body, the Baker-Ericksen  (BE) inequalities  state that  \emph{the greater principal Cauchy stress occurs in the direction of the greater principal stretch} \cite{BakerEricksen:1954}. In particular, under uniaxial tension, the deformation is a simple extension in the direction of the tensile force if and only if the BE inequalities hold \cite{Marzano:1983}. Under these mechanical constraints, the shear modulus of the material is positive \cite{Mihai:2017:MG}.

\item[(A4)] Finite mean and variance for the random shear modulus: we assume that, at any given finite deformation, the random shear modulus, $\mu$, and its inverse, $1/\mu$, are second-order random variables, i.e., they have finite mean value and finite variance \cite{Staber:2015:SG,Staber:2016:SG,Staber:2017:SG}.
\end{itemize}

Assumptions (A1)-(A3) are long-standing principles in isotropic finite elasticity \cite{goriely17,Ogden:1997,TruesdellNoll:2004}, while (A4) concerns physically realistic expectations on the random shear modulus, which will be described by a probability distribution.

In particular, we confine our attention to a class of stochastic homogeneous incompressible hyperelastic materials described by the constitutive law \cite{Staber:2017:SG,Mihai:2018a:MWG},
\begin{equation}\label{eq:W:stoch}
\mathcal{W}(\lambda_{1},\lambda_{2},\lambda_{3})=\frac{\mu_{1}}{2m^2}\left(\lambda_{1}^{2m}+\lambda_{2}^{2m}+\lambda_{3}^{2m}-3\right)
+\frac{\mu_{2}}{2n^2}\left(\lambda_{1}^{2n}+\lambda_{2}^{2n}+\lambda_{3}^{2n}-3\right),
\end{equation}
where $m$ and $n$ are deterministic constants, and $\mu_{1}$ and $\mu_{2}$ are random variables. Consistent with the deterministic elastic theory \cite{Mihai:2017:MG}, the random shear modulus for infinitesimal deformations of  the stochastic model \eqref{eq:W:stoch} is defined as $\mu=\mu_{1}+\mu_{2}$. One can also consider models where the exponents $m$ and $n$ are characterised as random variables as well. However, this additional complexity is not relevant for the problem at hand.

For these materials, condition (A4) is guaranteed by the following constraints on the expected values \cite{Staber:2015:SG,Staber:2016:SG,Staber:2017:SG,Mihai:2018a:MWG,Mihai:2018b:MWG}:
\begin{eqnarray}\label{eq:Emu1}\begin{cases}
E\left[\mu\right]=\underline{\mu}>0,&\\
E\left[\log\ \mu\right]=\nu,& \mbox{such that $|\nu|<+\infty$},\label{eq:Emu2}\end{cases}
\end{eqnarray}
i.e., the mean value $\underline{\mu}$ of the shear modulus, $\mu$, is fixed and greater than zero, and the mean value of $\log\ \mu$ is fixed and finite, implying that both $\mu$ and $1/\mu$ are second-order random variables, i.e., they have finite mean and finite variance \cite{Soize:2000,Soize:2001}. These expected values are then used to find the maximum likelihood probability for the random shear modulus, $\mu$, with mean value $\underline{\mu}$, and standard deviation $\|\mu\|=\sqrt{\text{Var}[\mu]}$, defined as the square root of the variance, $\text{Var}[\mu]$. Critically, under the constraints (\ref{eq:Emu1}), $\mu$ follows a Gamma probability distribution with hyperparameters $\rho_{1}>0$ and $\rho_{2}>0$ satisfying
\begin{equation}\label{eq:rho12}
\underline{\mu}=\rho_{1}\rho_{2},\qquad
\|\mu\|=\sqrt{\rho_{1}}\rho_{2}.
\end{equation}
The corresponding probability density function takes the form \cite{Abramowitz:1964,Johnson:1994:JKB}
\begin{equation}\label{eq:mu:gamma}
g(\mu;\rho_{1},\rho_{2})=\frac{\mu^{\rho_{1}-1}e^{-\mu/\rho_{2}}}{\rho_{2}^{\rho_{1}}\Gamma(\rho_{1})},\qquad\mbox{for}\ \mu>0\ \mbox{and}\ \rho_{1}, \rho_{2}>0,
\end{equation}
where $\Gamma:\mathbb{R}^{*}_{+}\to\mathbb{R}$ is the complete Gamma function
\begin{equation}\label{eq:gamma}
\Gamma(z)=\int_{0}^{+\infty}t^{z-1}e^{-t}\text dt.
\end{equation}
When $\mu_{1}>0$ and $\mu_{2}>0$, we can define the auxiliary random variable \cite{Mihai:2018a:MWG}
\begin{equation}\label{eq:R12:b}
R_{1}=\frac{\mu_{1}}{\mu},
\end{equation}
such that $0<R_{1}<1$. Then, under the following constraints \cite{Staber:2015:SG,Staber:2016:SG,Staber:2017:SG,Mihai:2018a:MWG},
\begin{eqnarray}\begin{cases}
E\left[\log\ R_{1}\right]=\nu_{1},& \mbox{such that $|\nu_{1}|<+\infty$},\label{eq:ER1}\\
E\left[\log(1-R_{1})\right]=\nu_{2},& \mbox{such that $|\nu_{2}|<+\infty$},\label{eq:ER2}\end{cases}
\end{eqnarray}
the random variable $R_{1}$ follows a standard Beta distribution \cite{Abramowitz:1964,Johnson:1994:JKB}, with hyperparameters $\xi_{1}>0$ and $\xi_{2}>0$ satisfying
\begin{equation}\label{eq:xi12}
\underline{R}_{1}=\frac{\xi_{1}}{\xi_{1}+\xi_{2}},\qquad
\|R_{1}\|=\frac{1}{\xi_{1}+\xi_{2}}\sqrt{\frac{\xi_{1}\xi_{2}}{\xi_{1}+\xi_{2}+1}}.
\end{equation}
where $\underline{R}_{1}$ is the mean value, $\|R_{1}\|=\sqrt{\text{Var}[R_{1}]}$ is the standard deviation, and $\text{Var}[R_{1}]$ is the variance  of $R_{1}$. The corresponding probability density function takes the form
\begin{equation}\label{eq:betar}
\beta(r;\xi_{1},\xi_{2})=\frac{r^{\xi_{1}-1}(1-r)^{\xi_{2}-1}}{B(\xi_{1},\xi_{2})},\qquad \qquad\mbox{for}\ r\in(0,1)\ \mbox{and}\ \xi_{1}, \xi_{2}>0,
\end{equation}
where $B:\mathbb{R}^{*}_{+}\times\mathbb{R}^{*}_{+}\to\mathbb{R}$ is the Beta function
\begin{equation}\label{eq:beta}
B(x,y)=\int_{0}^{1}t^{x-1}(1-t)^{y-1}dt.
\end{equation}
Then, for the random coefficients $\mu_{1}=R_{1}\mu$ and $\mu_{2}=\mu-\mu_{1}$, the corresponding mean values are
\begin{equation}\label{eq:mu12:mean}
\underline{\mu}_{1}=\frac{\rho_{1}\rho_{2}\xi_{1}}{\xi_{1}+\xi_{2}},\qquad
\underline{\mu}_{2}=\frac{\rho_{1}\rho_{2}\xi_{2}}{\xi_{1}+\xi_{2}},
\end{equation}
and the variances and covariance are, respectively,
\begin{eqnarray}
&&\text{Var}\left[\mu_{1}\right]
=\frac{\rho_{1}\rho_{2}^2\xi_{1}\left(\xi_{1}^2+\xi_{1}+\xi_{2}+\xi_{1}\xi_{2}+\rho_{1}\xi_{2}\right)}{\left(\xi_{1}+\xi_{2}\right)^2\left(\xi_{1}+\xi_{2}+1\right)},\\
&&\text{Var}\left[\mu_{2}\right]
=\frac{\rho_{1}\rho_{2}^2\xi_{2}\left(\xi_{2}^2+\xi_{1}+\xi_{2}+\xi_{1}\xi_{2}+\rho_{1}\xi_{1}\right)}{\left(\xi_{1}+\xi_{2}\right)^2\left(\xi_{1}+\xi_{2}+1\right)},\\
&&\text{Cov}\left[\mu_{1},\mu_{2}\right]
=\frac{\rho_{1}\rho_{2}^2\xi_{1}\xi_{2}\left(\xi_{1}+\xi_{2}-\rho_{1}\right)}{\left(\xi_{1}+\xi_{2}\right)^2\left(\xi_{1}+\xi_{2}+1\right)}.
\end{eqnarray}

We further note that, when $\rho_{1}\to\infty$, assuming that the standard deviation, $\|\mu\|$, is constant, by \eqref{eq:rho12}, $\rho_2=\|\mu\|/\sqrt{\rho_1}$. Next, defining $u=\mu+\|\mu\|/\sqrt{\rho_{1}}>\|\mu\|/\sqrt{\rho_{1}}$, the probability density function \eqref{eq:mu:gamma} takes the form
\[
g_{1}(u-\|\mu\|/\sqrt{\rho_{1}};\rho_{1},\|\mu\|/\sqrt{\rho_{1}})=\frac{\left(u-\|\mu\|/\sqrt{\rho_{1}}\right)^{\rho_{1}-1}e^{-\left(u-\|\mu\|/\sqrt{\rho_{1}}\right)/\left(\|\mu\|/\sqrt{\rho_{1}}\right)}}{\left(\|\mu\|/\sqrt{\rho_{1}}\right)^{\rho_{1}}\Gamma(\rho_{1})}.
\]
Then, the limit of the above function as $\rho_{1}\to\infty$ is equal to
\begin{equation}\label{eq:mu:gnormal}
\lim_{\rho_{1}\to\infty}g_{1}(u-\|\mu\|/\sqrt{\rho_{1}};\rho_{1},\|\mu\|/\sqrt{\rho_{1}})=\frac{e^{-(u-\underline{\mu})^2/(2\|\mu\|^2)}}{\sqrt{2\pi}\|\mu\|}.
\end{equation}
Hence, the Gamma probability density function \eqref{eq:mu:gamma} is approximated by a normal (Gaussian) density function
\begin{equation}\label{eq:mu:normal}
h(u;\underline{\mu},\|\mu\|)=\frac{e^{-(u-\underline{\mu})^2/(2\|\mu\|^2)}}{\sqrt{2\pi}\|\mu\|},
\end{equation}
where $u$ is a random normal variable with mean value $\underline{\mu}$ and standard deviation $\|\mu\|$. 

When $\rho_1\approx 1$, the probability distribution \eqref{eq:mu:gamma} reduces to an exponential distribution,
\begin{equation}\label{eq:mu:epdf}
g_{2}(\mu;\rho_{2})=\frac{e^{-\mu/\rho_{2}}}{\rho_{2}},\qquad\mbox{for}\ \mu>0\ \mbox{and}\ \rho_{2}>0.
\end{equation}
In this case, the mean value, $\underline{\mu},$ and standard deviation, $\|\mu\|$, defined by \eqref{eq:rho12}, take comparable values. This situation may arise, for example, when the sampled data contain a lot of noise.

\subsection{Rubberlike material}

\begin{figure}[htbp]
	\begin{center}
		\includegraphics[width=1\textwidth]{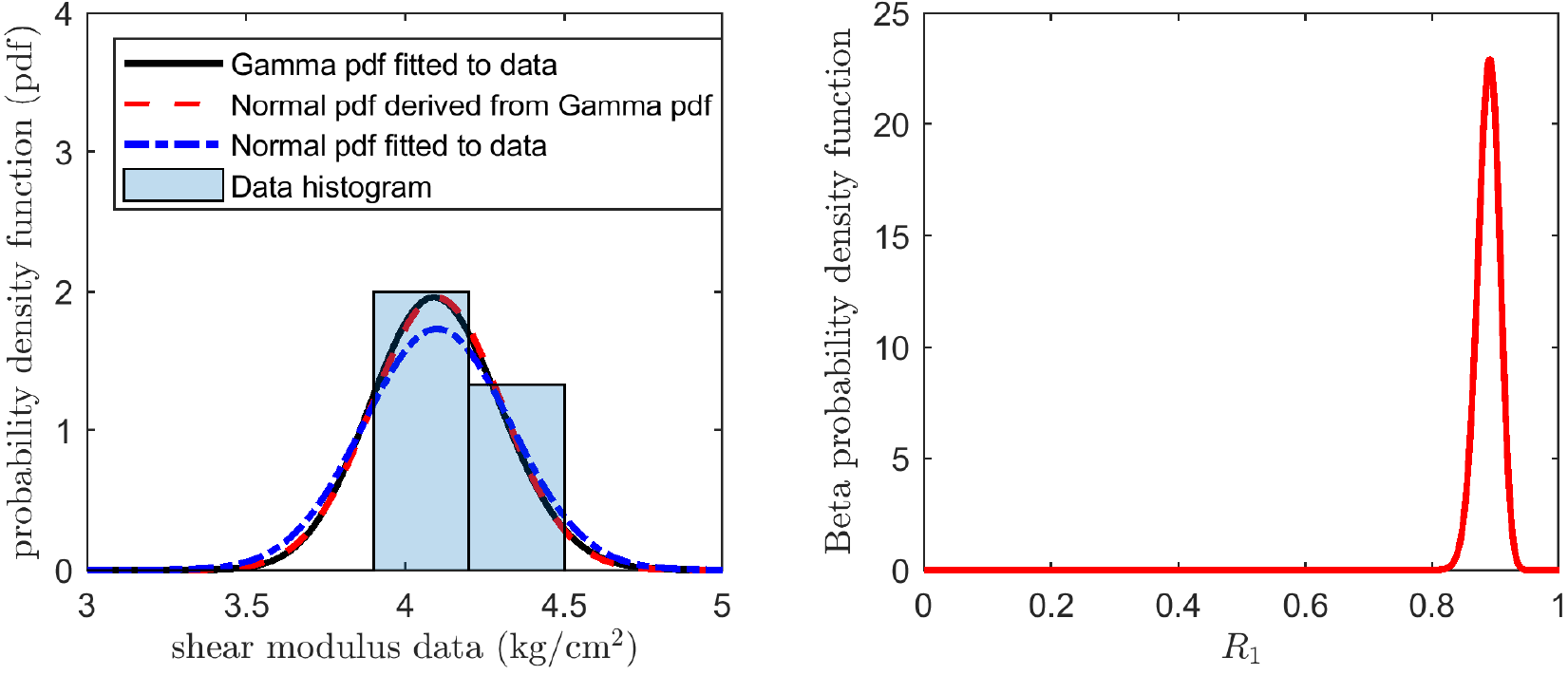}
		\caption{Probability distributions derived from the data values for the random shear modulus, $\mu=\mu_{1}+\mu_{2}$, given in Table~\ref{table1}. Left: the Gamma distribution takes the form \eqref{eq:mu:gamma}, while the normal distribution derived from it and the one fitted to data are given by \eqref{eq:mu:gnormal} and \eqref{eq:mu:normal}, respectively. The parameters for these  distributions are recorded in Table~\ref{table2}. Right: the Beta distribution takes the form \eqref{eq:betar}, with parameters recorded in Table~\ref{table3}.}\label{fig:rubberpdf}
	\end{center}
\end{figure}

For rubberlike material, the first experimental data in large deformations were reported by Rivlin and Saunders (1951) \cite{Rivlin:1951:VII:RS}. In light of these data, and assuming that such a material can be described by the stochastic hyperelastic model \eqref{eq:W:stoch} under sufficiently small deformations, here, we derive the probability distribution for the random shear modulus, $\mu=\mu_{1}+\mu_{2}$, for this material (see Figure~\ref{fig:rubberpdf}). Note that we only make this assumption in order to provide examples of probability distributions based on some real data measurements, and do not attempt to optimise a specific hyperelastic strain-energy function to the given data. For example, deterministic hyperelastic models calibrated to mean data values for rubberlike material under finite deformations were proposed in \cite{Destrade:2017:DSS,Hartmann:2001,Mihai:2017:MG,Ogden:2004:OSS,Steinmann:2012:etal,Twizell:1983:TO}, statistical models were derived computationally from artificially generated data in \cite{Caylak:2018:etal,Norenberg:2015:NM}, while explicit stochastic hyperelastic models based on available data sets consisting of mean values and standard deviations were obtained in \cite{Mihai:2018a:MWG}.

The chosen data values are recorded in Table~\ref{table1}. The Gamma probability distribution fitted to the shear modulus data, together with the normal distribution derived from the Gamma distribution, and also the standard normal distribution fitted to the data, are represented in Figure~\ref{fig:rubberpdf}. Note the similarity between the Gamma and normal distributions in this case. For each probability distribution, the mean value, $\underline{\mu}$, and standard deviation, $\|\mu\|$, are recorded in Table~\ref{table2}. For the normal distribution derived from the Gamma distribution, the mean value and standard deviation are given by \eqref{eq:rho12}, as for the Gamma distribution. For the auxiliary random variable $R_{1}=\mu_{1}/\mu$, the mean value, $\underline{R_{1}}$, and standard deviation, $\|R_{1}\|$, are provided in Table~\ref{table3}.

\begin{table}[h!!!t!!!b!!!p]
	\centering
	\renewcommand{\arraystretch}{2}
	\caption{Experimental data for rubberlike material under sufficiently small deformations, with the values of $\mu_{1}/2$ and $\mu_{2}/2$ selected from Tables~1 and 2 of \cite{Rivlin:1951:VII:RS}.}\label{table1}
	\vspace*{0.15cm}
	\scalebox{0.95}{
		\begin{tabular}{c|ccccc}
			$\lambda_{1}$ & 1.90 & 1.80 & 1.70 & 1.90 & 1.80 \\
			\hline
			$\lambda_{2}$ & 1.07 & 1.25 & 1.39 & 1.02 & 1.09 \\
			\hline
			$\mu_{1}/2$ (kg/cm$^2$) & 1.77 & 1.89 & 2.01 & 1.68 & 1.76  \\
			\hline
			$\mu_{2}/2$ (kg/cm$^2$) & 0.20 & 0.23 & 0.21 & 0.29 & 0.21 \\
			\hline
			$\mu=\mu_{1}+\mu_{2}$ (kg/cm$^2$) & 3.94 & 4.24 & 4.44 & 3.94 & 3.94 \\
		\end{tabular}
	}
	\centering
	\renewcommand{\arraystretch}{2}
	\caption{Parameters of the probability distributions derived from the data values for the random shear modulus, $\mu=\mu_{1}+\mu_{2}$, given in Table~\ref{table1}.}\label{table2}
	\vspace*{0.15cm}
	\scalebox{0.95}{
		\begin{tabular}{c|cccc}
			Probability density function (pdf) & $\underline{\mu}$ & $\|\mu\|$ & $\rho_1$ & $\rho_2$\\
			\hline
			Gamma pdf \eqref{eq:mu:gamma} fitted to data & $4.0907$ & $0.2037$ & $405.0214$ & $0.0101$\\
			\hline
			Normal pdf \eqref{eq:mu:gnormal} derived from Gamma pdf & $4.0907$ & $0.2037$ & - & -\\
			\hline
			Normal pdf \eqref{eq:mu:normal} fitted to data & $4.0907$ & $0.2302$ & - & -\\
		\end{tabular}
	}
\centering
\renewcommand{\arraystretch}{2}
\caption{Parameters of the probability distribution for the random variable $R_{1}=\mu_{1}/\mu$ derived from the data values provided in Table~\ref{table1}.}\label{table3}
\vspace*{0.15cm}
\scalebox{0.95}{
	\begin{tabular}{c|cccc}
		Probability density function (pdf) & $\underline{R}_{1}$ & $\|R_{1}\|$ & $\xi_1$ & $\xi_2$\\
		\hline
		Beta pdf fitted to data & $0.8883$ & $0.0175$ & $287.2297$ & $36.1194$\\
	\end{tabular}
}
\end{table}

In the next sections, we analyse the radially symmetric inflation of a spherical shell and of a cylindrical tube of stochastic hyperelastic material defined by  \eqref{eq:W:stoch}, with the random shear modulus, $\mu$, following either the Gamma or the normal probability distribution fitted to the given data. We can regard the stochastic spherical shell (or tube) as an ensemble (or population) of shells (tubes) with the same geometry, such that each shell (tube) is made from a homogeneous isotropic incompressible hyperelastic material, with the elastic parameters not known with certainty, but drawn from known probability distributions. Then, for each individual shell (tube), the finite elasticity theory applies. The question is: \emph{what is the probability distribution of stable radially symmetric inflation, such that the internal pressure always increases as the radial stretch increases?}

\section{Stochastic incompressible spherical shell}\label{sec:shell}

\begin{figure}[htbp]
	\begin{center}
		\includegraphics[width=0.7\textwidth]{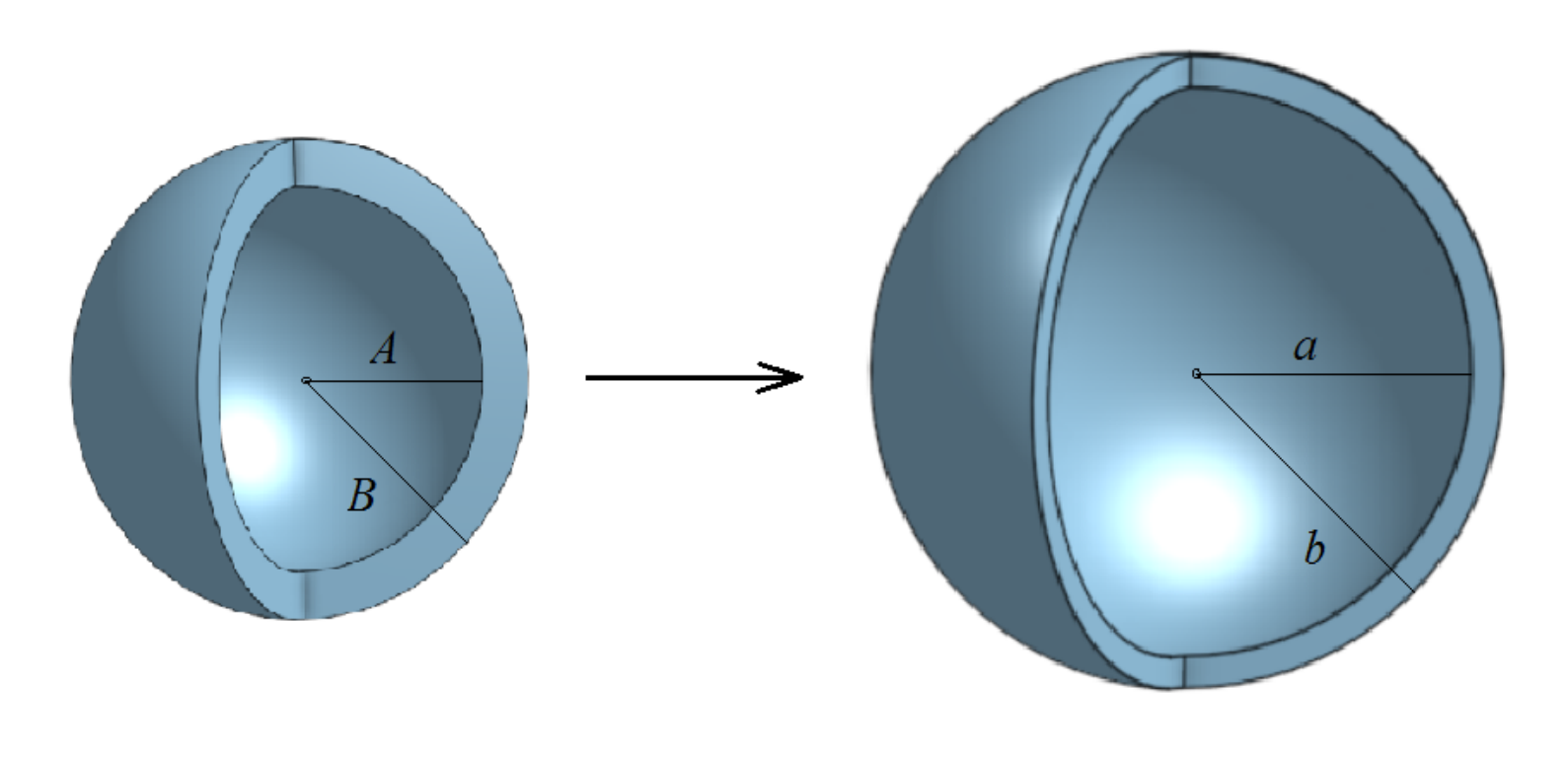}
		\caption{Schematic of inflation of a spherical shell, showing the reference state, with inner radius $A$ and outer radius $B$ (left), and the deformed state, with inner radius $a$ and outer radius $b$ (right), respectively.}\label{fig:shell1}
	\end{center}
	\begin{center}
        \includegraphics[width=0.6\textwidth]{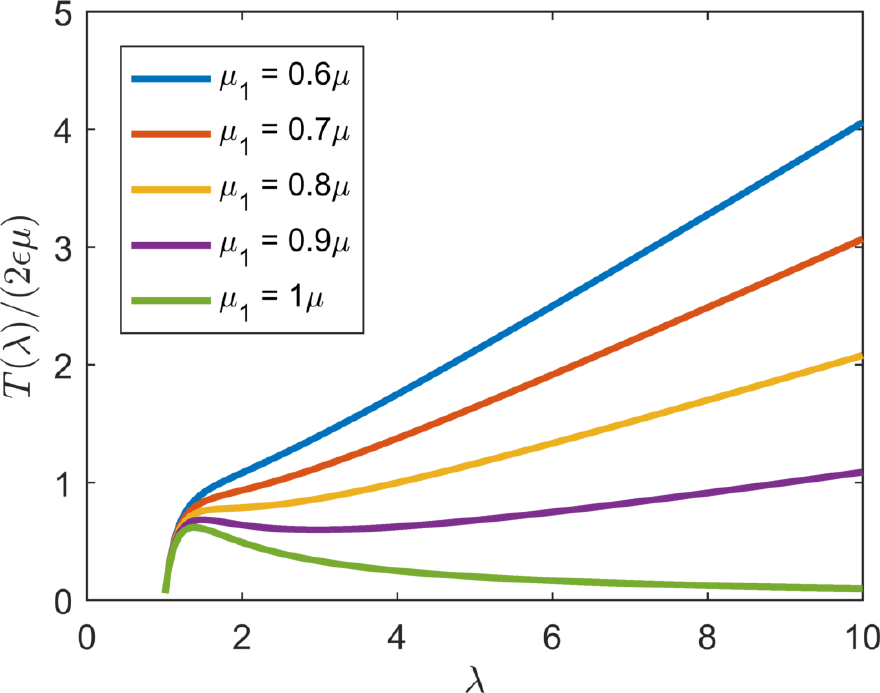}
		\caption{The normalised internal pressure, $T(\lambda)$ defined by \eqref{eq:shell:T:0}, for the inflation of spherical shells of Mooney-Rivlin materials, with $m=1$ and $n=-1$. In this deterministic case, inflation instability occurs if $\mu_{1}/\mu>0.8234$.}\label{fig:shell2}
	\end{center}
\end{figure}

We consider first a spherical shell of stochastic hyperelastic material described by \eqref{eq:W:stoch}, subject to the following radially symmetric deformation (see Figure~\ref{fig:shell1}),
\begin{equation}\label{eq:shell:deform}
r=f(R)R,\qquad \theta=\Theta,\qquad \phi=\Phi,
\end{equation}
where $(R,\Theta,\Phi)$ and $(r,\theta,\phi)$ are the spherical polar coordinates in the reference and the current configuration, respectively, such that $A\leq R\leq B$, and $f(R)\geq 0$ is to be determined.

The deformation gradient is $\textbf{F}=\text{diag}\left(\lambda_{1},\lambda_{2},\lambda_{3}\right)$, with
\begin{equation}\label{eq:shell:lambdas}
\lambda_{1}=f(R)+R\frac{\text{d}f}{\text{d}R}=\lambda^{-2},\qquad \lambda_{2}=\lambda_{3}=f(R)=\lambda,
\end{equation}
where $\lambda_{1}$, $\lambda_{2}$, $\lambda_{3}$ represent the radial, tangential, and azimuthal stretch ratios, respectively, and $\text{d}f/\text{d}R$ denotes differentiation of $f$ with respect to $R$.

The radial equation of equilibrium is \cite{Carroll:1987}
\begin{equation}\label{eq:equilib}
\frac{\text{d}P_{11}}{\text{d}R}+\frac{2}{R}(P_{11}-P_{22})=0,
\end{equation}
or equivalently,
\begin{equation}\label{eq:shell:equilib}
\frac{\text{d}P_{11}}{\text{d}\lambda}\lambda^{-2}+2\frac{P_{11}-P_{22}}{1-\lambda^{3}}=0,
\end{equation}
where $\textbf{P}=(P_{ij})_{i,j=1,2,3}$ is the first Piola-Kirchhoff stress tensor. For an incompressible material,
\begin{equation}\label{eq:shell:P11P22}
P_{11}=\frac{\partial\mathcal{W}}{\partial\lambda_{1}}-\frac{p}{\lambda_{1}},\qquad
P_{22}=\frac{\partial\mathcal{W}}{\partial\lambda_{2}}-\frac{p}{\lambda_{2}},
\end{equation}
where $p$ is the Lagrange multiplier for the incompressiblity constraint ($\det\textbf{F}=1$).

\subsection{Limit-point instability criterion for spherical shells}

Denoting
\begin{equation}\label{eq:shell:W}
W(\lambda)=\mathcal{W}(\lambda^{-2},\lambda,\lambda),
\end{equation}
where $\lambda=r/R>1$, we obtain
\begin{equation}\label{eq:shell:dW}
\frac{\text{d}W}{\text{d}\lambda}=-\frac{2}{\lambda^3}\frac{\partial\mathcal{W}}{\partial\lambda_{1}}+2\frac{\partial\mathcal{W}}{\partial\lambda_{2}}
=-\frac{2P_{11}}{\lambda^3}+2P_{22}.
\end{equation}
Next, setting the external pressure (at $R=B$) equal to zero, by \eqref{eq:shell:equilib} and \eqref{eq:shell:dW}, the internal pressure (at $R=A$) is equal to
\begin{equation}\label{eq:shell:integral:T}
\begin{split}
T&=-\frac{P_{11}}{\lambda^2}\left|_{\lambda=\lambda_{a}}\right.,\\
&=-2\int_{\lambda_{a}}^{\lambda_{b}}\frac{P_{11}}{\lambda^3}\text{d}\lambda+\int_{\lambda_{a}}^{\lambda_{b}}\frac{dP_{11}}{\text{d}\lambda}\lambda^{-2}\text{d}\lambda,\\
&=-2\int_{\lambda_{a}}^{\lambda_{b}}\frac{P_{11}}{\lambda^3}\text{d}\lambda-2\int_{\lambda_{a}}^{\lambda_{b}}\frac{P_{11}-P_{22}}{1-\lambda^3}\text{d}\lambda,\\
&=\int_{\lambda_{a}}^{\lambda_{b}}\frac{\text{d}W}{\text{d}\lambda}\frac{\text{d}\lambda}{1-\lambda^3},
\end{split}
\end{equation}
where $\lambda_{a}=a/A$ and $\lambda_{b}=b/B$ are the stretch ratios for the inner and outer radii, respectively. We recall that a volume element $\text{d}V$ from the reference configuration is transformed, after the deformation, into a volume element $\text{d}v=(\det\textbf{F})\text{d}V$ in the current configuration \cite[p.~240]{TruesdellNoll:2004}, \cite[p.~87]{Ogden:1997}, \cite[p.~274]{goriely17}. Then, by the material incompressibility condition, $\det\textbf{F}=1$, the material volume in the spherical shell is conserved, i.e., $4\pi\left(b^3-a^3\right)=4\pi\left(B^3-A^3\right)$, or equivalently, as $a=A\lambda_{a}$ and $b=B\lambda_{b}$,
\begin{equation}\label{eq:shell:lambdab}
\lambda_{b}^3=\left(\lambda_{a}^3-1\right)\left(\frac{A}{B}\right)^3+1.
\end{equation}
Hence, the internal pressure $T$ given by \eqref{eq:shell:integral:T} can be expressed as a function of the inner stretch ratio, $\lambda_{a}$, only.

As for the deterministic elastic shell, for the stochastic spherical shell, a \emph{limit-point instability} occurs if there is a change in the monotonicity of $T$, defined by \eqref{eq:shell:integral:T}, as a function of $\lambda_{a}$. When the spherical shell is thin, i.e., $0<\varepsilon=(B-A)/A\ll 1$, we can approximate the internal pressure as follows \cite[p.~443]{goriely17},
\begin{equation}\label{eq:shell:T}
T(\lambda)=\frac{\varepsilon}{\lambda^2}\frac{\text{d}W}{\text{d}\lambda},
\end{equation}
and find the critical value of $\lambda$ where a limit-point instability occurs by solving for $\lambda>1$ the following equation,
\begin{equation}\label{eq:shell:lps}
\frac{\text{d}T}{\text{d}\lambda}=0,
\end{equation}
where $T$ is described by \eqref{eq:shell:T}.

\begin{figure}[htbp]
	\begin{center}
		\includegraphics[width=\textwidth]{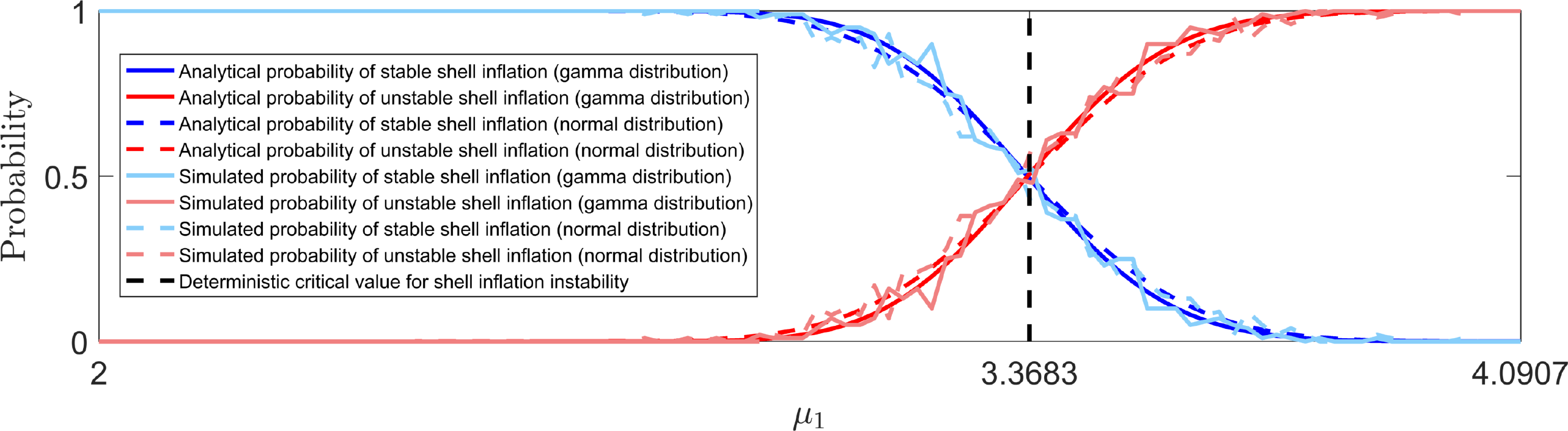}
		\caption{Probability distributions \eqref{eq:shell:P1}-\eqref{eq:shell:P2} of whether instability can occur or not for a spherical shell of stochastic Mooney-Rivlin material, described by \eqref{eq:W:stoch} with $m=1$ and $n=-1$, and the shear modulus, $\mu$, following either a Gamma distribution \eqref{eq:mu:gamma} with $\rho_{1}=405.0214$, $\rho_{2}=0.0101$ (continuous lines), or a normal distribution \eqref{eq:mu:normal} with $\underline{\mu}=4.0907$, $\|\mu\|=0.2302$ (dashed lines). Darker colours represent analytically derived solutions, given by equations \eqref{eq:shell:P1}-\eqref{eq:shell:P2}, whereas lighter colours represent stochastically generated data. The vertical line at the critical value, $\mu_{1}=3.3683$, separates the expected regions based only on the mean value of the shear modulus. }\label{fig:shell3}
	\end{center}
	\begin{center}
		\includegraphics[width=0.6\textwidth]{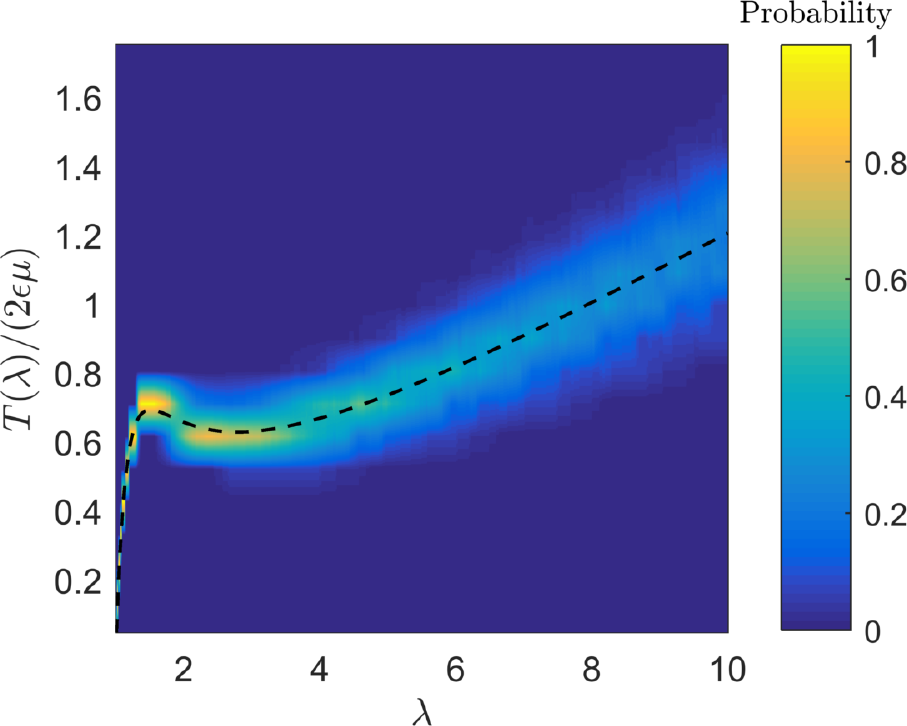}
		\caption{Computed probability distribution of the normalised internal pressure, $T(\lambda)$ defined by \eqref{eq:shell:T}, for the inflation of a spherical shell of stochastic Mooney-Rivlin material, given by  \eqref{eq:W:stoch} with $m=1$ and $n=-1$, when $\mu$ follows a Gamma distribution \eqref{eq:mu:gamma} with $\rho_{1}=405.0214$, $\rho_{2}=0.0101$, and $R_{1}=\mu_{1}/\mu$ follows a Beta distribution \eqref{eq:betar} with $\xi_{1}=287.2297$, $\xi_{2}=36.1194$. As $\underline{\mu}_{1}=3.6338=0.8883\cdot\underline{\mu}>0.8234\cdot\underline{\mu}$, instability is expected to occur, but there is also around 10\% chance that the inflation is stable. The dashed black line corresponds to the expected pressure based only on mean parameter values.}\label{fig:shell4}
	\end{center}
\end{figure}

\subsection{Deterministic elastic shell}

In the deterministic case, for a spherical shell of hyperelastic material defined by the strain-energy function taking the form \eqref{eq:W:stoch}, but with $\mu_{1}$ and $\mu_{2}$ fixed positive constants, and $\mu=\mu_{1}+\mu_{2}>0$ the corresponding shear modulus, the function \eqref{eq:shell:W} is equal to
\begin{equation}\label{eq:shell:W:0}
W(\lambda)=\frac{\mu_{1}}{2m^2}\left(\lambda^{-4m}+2\lambda^{2m}-3\right)+\frac{\mu_{2}}{2n^2}\left(\lambda^{-4n}+2\lambda^{2n}-3\right).
\end{equation}
Then, the  internal pressure given by \eqref{eq:shell:T} takes the form
\begin{equation}\label{eq:shell:T:0}
T(\lambda)=2\varepsilon\left[\frac{\mu_{1}}{m}\left(\lambda^{2m-3}-\lambda^{-4m-3}\right)+\frac{\mu_{2}}{n}\left(\lambda^{2n-3}-\lambda^{-4n-3}\right)\right],
\end{equation}
and the equation \eqref{eq:shell:lps} is equivalent to
\begin{equation}\label{eq:shell:dT:0}
\frac{\mu_{1}}{m}\left[(2m-3)\lambda^{2m-4}+(4m+3)\lambda^{-4m-4}\right]+\frac{\mu_{2}}{n}\left[(2n-3)\lambda^{2n-4}+(4n+3)\lambda^{-4n-4}\right]=0.
\end{equation}
Specifically, if $\mu_{2}=0$,  then, when $-3/4<m<3/2$, the internal pressure increases to a maximum value then decreases, otherwise the internal pressure increases monotonically. If $\mu_{2}>0$, then \eqref{eq:shell:dT:0} is equivalent to
\begin{equation}\label{eq:shell:mu1}
\frac{\mu_{1}}{\mu}=\frac{m\left[(2n-3)\lambda^{2n-4}+(4n+3)\lambda^{-4n-4}\right]}{m\left[(2n-3)\lambda^{2n-4}+(4n+3)\lambda^{-4n-4}\right]-n\left[(2m-3)\lambda^{2m-4}+(4m+3)\lambda^{-4m-4}\right]},
\end{equation}
where $0< \mu_{1}/\mu<1$.

In particular, for a spherical shell of Mooney-Rivlin material, with $m=1$ and $n=-1$, the minimum value of $\mu_{1}/\mu$, such that inflation instability occurs, is the minimum value of the following function,
\begin{equation}\label{eq:shell:h}
\eta(\lambda)=\frac{\lambda^8+5\lambda^2}{\lambda^8+5\lambda^2+\lambda^6-7}, \qquad \lambda\geq 7^{1/6},
\end{equation}
i.e., 
\begin{equation}\label{eq:shell:etamin}
\eta_{min}\approx 0.8234.
\end{equation}
In this case:
\begin{itemize}
	\item[(i)] When $0.8234\leq \mu_{1}/\mu< 1$, the internal pressure increases to a maximum, then decreases to a minimum, then increases again;
	\item[(ii)] When $0<\mu_{1}/\mu<0.8234$, the internal pressure is always increasing.
\end{itemize}

For spherical shells of Mooney-Rivlin material, with $m=1$ and $n=-1$, the internal pressure, $T(\lambda)$ defined by \eqref{eq:shell:T:0} and normalised by $2\varepsilon\mu$, is plotted in Figure~\ref{fig:shell2} (see also \cite{Carroll:1987,Goriely:2006:GDBA,Zamani:2017:ZP},  \cite[pp.~283-288]{Ogden:1997}, \cite[pp.~442-447]{goriely17} and the references therein).

\subsection{Stochastic elastic shell}

For a spherical shell of stochastic Mooney-Rivlin material, characterised by the strain-energy function \eqref{eq:W:stoch} with $m=1$ and $n=-1$, the probability distribution of stable inflation, such that internal pressure monotonically increases as the radial stretch increases, is
\begin{equation}\label{eq:shell:P1}
P_{1}(\mu_{1})=1-\int_{0}^{\mu_{1}/0.8234}p_{1}(u)du,
\end{equation}
where $0.8234$ is given by \eqref{eq:shell:etamin}, $p_{1}(u)=g(u;\rho_{1},\rho_{2})$ if the random shear modulus, $\mu$, follows the Gamma distribution \eqref{eq:mu:gamma}, with $\rho_{1}=405.0214$ and $\rho_{2}=0.0101$, and $p_{1}(u)=h(u;\underline{\mu},\|\mu\|)$ if $\mu$ follows the normal distribution \eqref{eq:mu:normal}, with $\underline{\mu}=4.0907$ and $\|\mu\|=0.2302$ (see table~\ref{table2}).

The probability distribution of inflation instability occurring, such that the internal pressure begins to decrease, is
\begin{equation}\label{eq:shell:P2}
P_{2}(\mu_{1})=1-P_{1}(\mu_{1}).
\end{equation}

The probability distributions given by equations \eqref{eq:shell:P1}-\eqref{eq:shell:P2} are illustrated numerically in Figure~\ref{fig:shell3} (blue lines for $P_{1}$ and red lines for $P_{2}$). Specifically, $\mu_{1}\in(0,\underline{\mu})$ was divided into $100$ steps, then for each value of $\mu_{1}$, $100$ random values of $\mu$ were numerically generated from a specified Gamma (or normal) distribution and compared with the inequalities defining the two intervals for values of $\mu_{1}$. For the deterministic elastic shell, which is based on the mean value of the shear modulus, $\underline{\mu}=\rho_{1}\rho_{2}=4.0907$, the critical value of $\mu_{1}=0.8234\cdot\mu=3.3683$ strictly separates the cases where inflation instability can occur or not. For the stochastic problem, for the same critical value, there is, by definition, exactly 50\% chance of a randomly chosen shell for which inflation is stable (blue solid or dashed line if the shear modulus is Gamma or normal distributed, respectively), and 50\% chance of a randomly chosen shell, such that a limit-point instability occurs (red solid or dashed line). To increase the probability of stable inflation ($P_{1}\approx 1$), one must consider sufficiently small values of $\mu_{1}$, below the expected critical value, whereas a limit-point instability is certain to occur ($P_{2}\approx 1$) only if the model reduces to the neo-Hookean case. However, the inherent variability in the probabilistic system means that there will also exist events where there is competition between the two cases.

To illustrate this, in Figure~\ref{fig:shell4}, we show the probability distribution of the normalised internal pressure $T(\lambda)$, defined by \eqref{eq:shell:T}, as a function of the inner stretch $\lambda$, when $\mu$ follows a Gamma distribution with $\rho_{1}=405.0214$ and $\rho_{2}=0.0101$, and $R_{1}=\mu_{1}/\mu$ follows a Beta distribution with $\xi_{1}=287.2297$ and $\xi_{2}=36.1194$  (see table~\ref{table3}). In this case, $\underline{\mu}_{1}=3.6338=0.8883\cdot\underline{\mu}>0.8234\cdot\underline{\mu}$, and instability is expected to occur. Nevertheless, the probability distribution suggests that there is also around 10\% chance that the inflation is stable.

\section{Stochastic incompressible cylindrical tube}\label{sec:tube}

\begin{figure}[htbp]
	\begin{center}
		\includegraphics[width=0.5\textwidth]{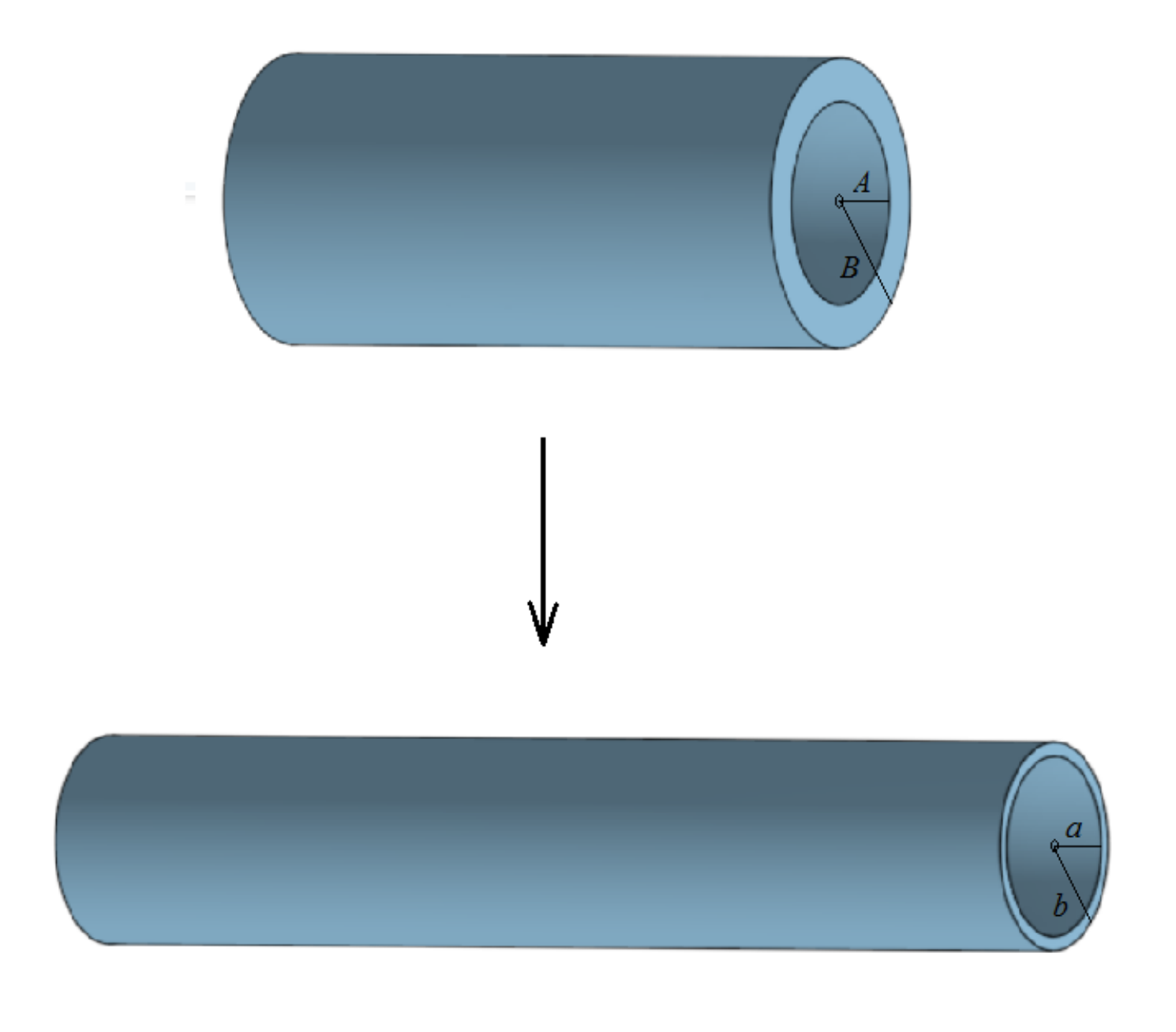}
		\caption{Schematic of inflation and stretching of a cylindrical tube, showing the reference state, with inner radius $A$ and outer radius $B$ (top), and the deformed state, with inner radius $a$ and outer radius $b$ (bottom), respectively.}\label{fig:tube1}
	\end{center}
	\begin{center}
		\includegraphics[width=0.6\textwidth]{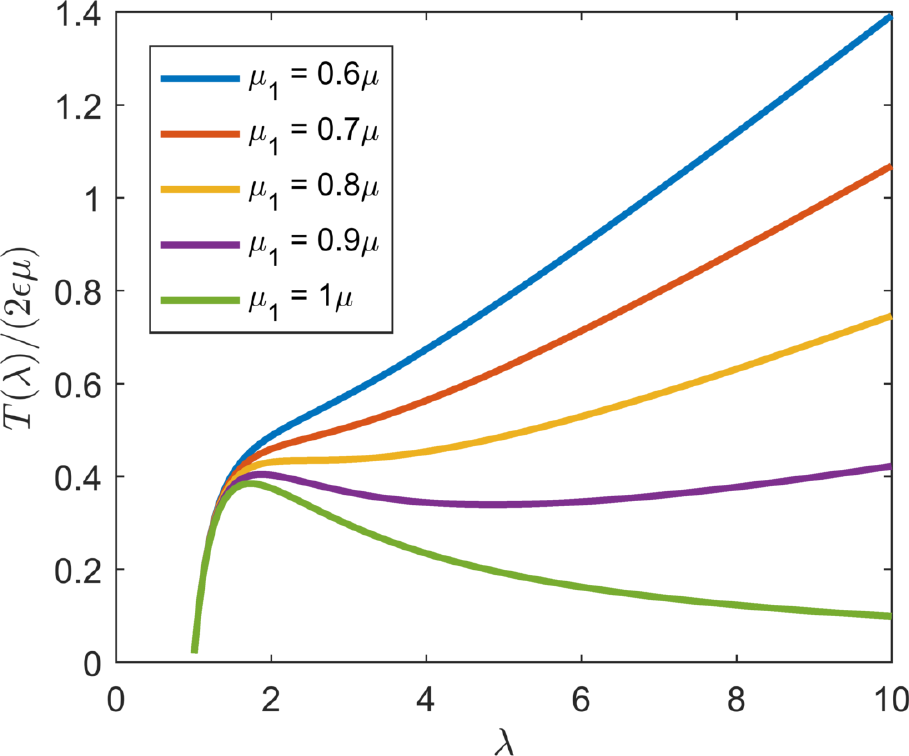}
		\caption{The normalised internal pressure, $T(\lambda)$ defined by \eqref{eq:tube:T:0}, for the inflation of cylindrical tubes of hyperelastic materials, with $m=1/2$ and $n=-3/2$. In this deterministic case, inflation instability occurs if $\mu_{1}/\mu>0.8035$.}\label{fig:tube2}
	\end{center}
\end{figure}

Next, a cylindrical tube of stochastic hyperelastic material given by \eqref{eq:W:stoch} is deformed through the combined effects of radially symmetric inflation and axial extension \cite{Rivlin:1949:VI}, as follows (see Figure~\ref{fig:tube1}),
\begin{equation}\label{eq:tube:deform}
r=f(R)R,\qquad \theta=\Theta,\qquad z=\alpha Z,
\end{equation}
where $(R,\Theta,Z)$ and $(r,\theta,z)$ are the cylindrical polar coordinates in the reference and the current configuration, respectively, such that $A\leq R\leq B$, $\alpha>0$ is a given constant (when $\alpha<0$, the tube is inverted, so that the inner surface becomes the outer surface), and $f(R)\geq 0$ is to be determined.

The deformation gradient is $\textbf{F}=\text{diag}\left(\lambda_{1},\lambda_{2},\lambda_{3}\right)$, with
\begin{equation}\label{eq:tube:lambdas}
\lambda_{1}=f(R)+R\frac{\text{d}f}{\text{d}R}=\lambda^{-1}\alpha^{-1},\qquad \lambda_{2}=f(R)=\lambda,\qquad \lambda_{3}=\alpha,
\end{equation}
where $\lambda_{1}$, $\lambda_{2}$, $\lambda_{3}$ are the radial, tangential, and longitudinal stretch ratios, respectively.

For the cylindrical tube, the radial equation of equilibrium is equivalent to
\begin{equation}\label{eq:tube:equilib}
\frac{\text{d}P_{11}}{\text{d}\lambda}\lambda^{-1}\alpha^{-1}+\frac{P_{11}-P_{22}}{1-\lambda^{2}\alpha}=0.
\end{equation}

\subsection{Limit-point instability criterion for cylindrical tubes}

Denoting
\begin{equation}\label{eq:tube:W}
W(\lambda)=\mathcal{W}(\lambda^{-1}\alpha^{-1},\lambda,\alpha),
\end{equation}
where $\lambda=r/R$ and $\alpha=z/Z$, we obtain
\begin{equation}\label{eq:tube:dW}
\frac{\text{d}W}{\text{d}\lambda}=-\frac{1}{\lambda^2\alpha}\frac{\partial\mathcal{W}}{\partial\lambda_{1}}+\frac{\partial\mathcal{W}}{\partial\lambda_{2}}
=-\frac{P_{11}}{\lambda^2\alpha}+P_{22}.
\end{equation}
Setting the external pressure (at $R=B$) equal to zero, by \eqref{eq:tube:equilib} and \eqref{eq:tube:dW}, the internal pressure (at $R=A$) is equal to
\begin{equation}\label{eq:tube:integral:T}
T=-\frac{P_{11}}{\lambda\alpha}\left|_{\lambda=\lambda_{a}}\right.=\int_{\lambda_{a}}^{\lambda_{b}}\frac{\text{d}W}{\text{d}\lambda}\frac{\text{d}\lambda}{1-\lambda^2\alpha},
\end{equation}
where $\lambda_{a}=a/A$ and $\lambda_{b}=b/B$ are the stretches for the inner and outer radii, respectively. Due to the material incompressibility, the material volume in the cylindrical tube is conserved, i.e., $\pi\alpha\left(b^2-a^2\right)=\pi\left(B^2-A^2\right)$, or equivalently, as $a=A\lambda_{a}$ and $b=B\lambda_{b}$,
\begin{equation}\label{eq:tube:lambdab}
\lambda_{b}^2=\left(\lambda_{a}^2-\frac{1}{\alpha}\right)\left(\frac{A}{B}\right)^2+\frac{1}{\alpha}.
\end{equation}
Hence, the internal pressure $T$ described by \eqref{eq:tube:integral:T} is a function of the inner stretch ratio, $\lambda_{a}$, only.

For the cylindrical tube, a \emph{limit-point instability} occurs if there is a change in the monotonicity of $T$, given by \eqref{eq:tube:integral:T}, as a function of $\lambda_{a}$. Assuming that the tube is thin, i.e., $0<\varepsilon=(B-A)/A\ll 1$, we approximate the internal pressure as  \cite[p.~290]{Ogden:1997}
\begin{equation}\label{eq:tube:T}
T(\lambda)= \frac{\varepsilon}{\lambda\alpha}\frac{\text{d}W}{\text{d}\lambda},
\end{equation}
and find the point of instability by solving for $\lambda>1$ the following equation,
\begin{equation}\label{eq:tube:lps}
\frac{\text{d}T}{\text{d}\lambda}=0,
\end{equation}
with $T$ given by \eqref{eq:tube:T}.

\begin{figure}[htbp]
	\begin{center}
		\includegraphics[width=\textwidth]{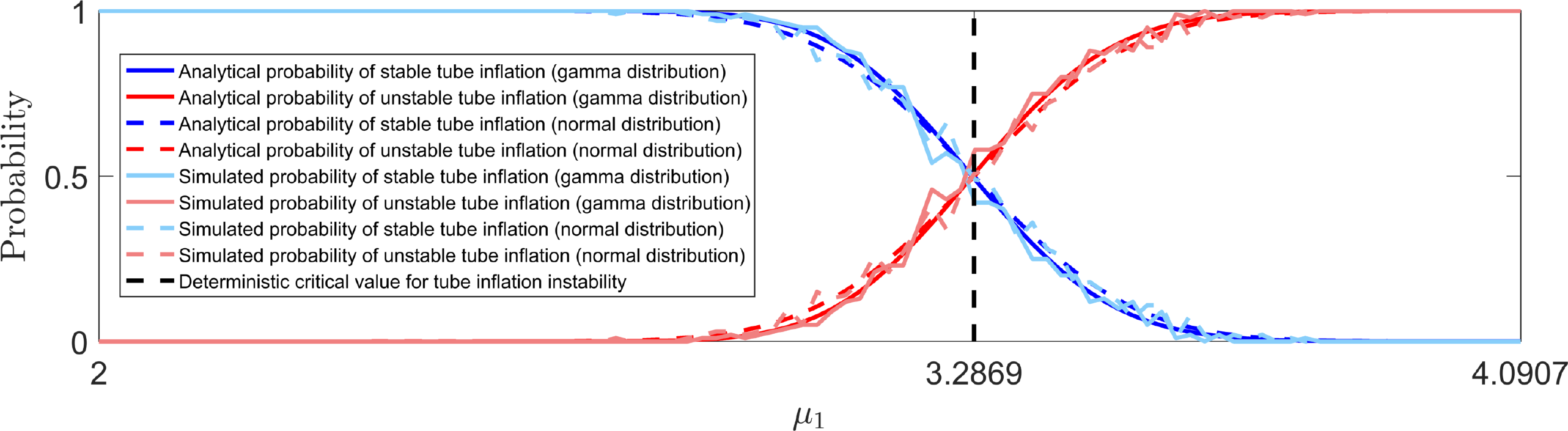}
		\caption{Probability distributions \eqref{eq:tube:P1}-\eqref{eq:tube:P2} of whether instability can occur or not for a cylindrical tube of stochastic hyperelastic material, described by \eqref{eq:W:stoch} with $m=1/2$ and $n=-3/2$, and the shear modulus, $\mu$, following either a Gamma distribution \eqref{eq:mu:gamma} with $\rho_{1}=405.0214$, $\rho_{2}=0.0101$ (continuous lines), or a normal distribution \eqref{eq:mu:normal} with $\underline{\mu}=4.0907$, $\|\mu\|=0.2302$ (dashed lines). Darker colours represent analytically derived solutions, given by equations \eqref{eq:tube:P1}-\eqref{eq:tube:P2}, whereas lighter colours represent stochastically generated data. The vertical line at the critical value, $\mu_{1}=3.2869$, separates the expected regions based only on the mean value of the shear modulus. }\label{fig:tube3}
	\end{center}
	\begin{center}
		\includegraphics[width=0.6\textwidth]{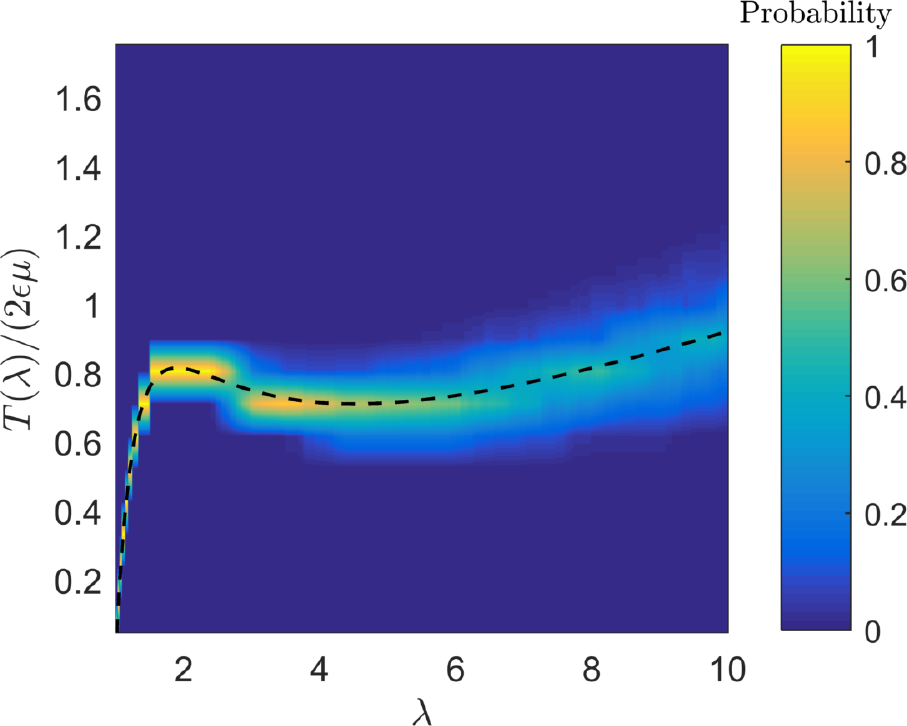}
		\caption{Computed probability distribution of the normalised internal pressure, $T(\lambda)$ defined by \eqref{eq:tube:T}, for the inflation of a cylindrical tube of stochastic hyperelastic material, given by  \eqref{eq:W:stoch} with $m=1/2$ and $n=-3/2$, when $\mu$ follows a Gamma distribution \eqref{eq:mu:gamma} with $\rho_{1}=405.0214$, $\rho_{2}=0.0101$, and $R_{1}=\mu_{1}/\mu$ follows a Beta distribution \eqref{eq:betar} with $\xi_{1}=287.2297$, $\xi_{2}=36.1194$. As $\underline{\mu}_{1}=3.6338=0.8883\cdot\underline{\mu}>0.8035\cdot\underline{\mu}$, instability is expected to occur, but there is also around 5\% chance that the inflation is stable. The dashed black line corresponds to the expected pressure based only on mean parameter values.}\label{fig:tube4}
	\end{center}
\end{figure}

\subsection{Deterministic elastic tube}

In the deterministic case, for a cylindrical tube of hyperelastic material defined by the strain-energy function \eqref{eq:W:stoch}, with $\mu_{1}$ and $\mu_{2}$ given positive constant, the corresponding function \eqref{eq:tube:W} takes the form
\begin{equation}\label{eq:tube:W:0}
W(\lambda)=\frac{\mu_{1}}{2m^2}\left(\lambda^{-2m}\alpha^{-2m}+\lambda^{2m}+\alpha^{2m}-3\right)+\frac{\mu_{2}}{2n^2}\left(\lambda^{-2n}\alpha^{-2n}+\lambda^{2n}+\alpha^{2n}-3\right).
\end{equation}
Then, the internal pressure given by \eqref{eq:tube:T} is equal to
\begin{equation}\label{eq:tube:T:0}
T(\lambda)=\frac{\varepsilon}{\alpha}\left[\frac{\mu_{1}}{m}\left(\lambda^{2m-2}-\lambda^{-2m-2}\alpha^{-2m}\right)+\frac{\mu_{2}}{n}\left(\lambda^{2n-2}-\lambda^{-2n-2}\alpha^{-2n}\right)\right],
\end{equation}
and the equation \eqref{eq:tube:lps} becomes
\begin{equation}\label{eq:tube:dT:0}
\frac{\mu_{1}}{m}\left[(m-1)\lambda^{2m-3}+(m+1)\lambda^{-2m-3}\alpha^{-2m}\right]+\frac{\mu_{2}}{n}\left[(n-1)\lambda^{2n-3}+(n+1)\lambda^{-2n-3}\alpha^{-2n}\right]=0.
\end{equation}
In this case, if $\mu_{2}=0$, then, when $-1<m<1$, the internal pressure increases to a maximum value then decreases, otherwise the internal pressure is always increasing. If $\mu_{2}>0$, then \eqref{eq:tube:dT:0} is equivalent to
\begin{equation}\label{eq:tube:mu1}
\frac{\mu_{1}}{\mu}=\frac{m\left[(n-1)\lambda^{2n-3}+(n+1)\lambda^{-2n-3}\alpha^{-2n}\right]}{m\left[(n-1)\lambda^{2n-3}+(n+1)\lambda^{-2n-3}\alpha^{-2n}\right]-n\left[(m-1)\lambda^{2m-3}+(m+1)\lambda^{-2m-3}\alpha^{-2m}\right]},
\end{equation}
where $0< \mu_{1}/\mu<1$.

Next, we specialise to the case where $m=1/2$, $n=-3/2$, and $\alpha=1$, for which \eqref{eq:tube:mu1} takes the form
\begin{equation}\label{eq:tube:mu1:ex}
\frac{\mu_{1}}{\mu}=\frac{1+5\lambda^{-6}}{1+5\lambda^{-6}+3\lambda^{-2}-9\lambda^{-4}}.
\end{equation}
Then, the minimum value of $\mu_{1}/\mu$, such that inflation instability occurs, is the minimum of the function
\[
\eta(\lambda)=\frac{\lambda^6+5}{\lambda^6+5+3\lambda^{4}-9\lambda^{2}}, \qquad \lambda> 1,
\]
i.e., 
\begin{equation}\label{eq:tube:etamin}
\eta_{min}\approx 0.8035.
\end{equation}

For cylindrical tubes of hyperelastic material with $m=1/2$ and $n=-3/2$, under the deformation \eqref{eq:tube:deform}, with $\alpha=1$, the internal pressure, $T(\lambda)$ defined by \eqref{eq:tube:T:0} and normalised by $2\varepsilon\mu$, is plotted in Figure~\ref{fig:tube2} (see also \cite{Carroll:1987}, \cite[pp.~288-291]{Ogden:1997}).

\subsection{Stochastic elastic tube}

For a cylindrical tube of stochastic hyperelastic material described by the strain-energy function \eqref{eq:W:stoch} with $m=1/2$ and $n=-3/2$, the probability distribution of stable inflation, such that the internal pressure always increases as the radial stretch increases, is
\begin{equation}\label{eq:tube:P1}
P_{1}(\mu_{1})=1-\int_{0}^{\mu_{1}/0.8035}p_{1}(u)du,
\end{equation}
where $0.8035$ is given by \eqref{eq:tube:etamin}, $p_{1}(u)=g(u;\rho_{1},\rho_{2})$ if the random shear modulus, $\mu$, follows the Gamma distribution \eqref{eq:mu:gamma}, with $\rho_{1}=405.0214$ and $\rho_{2}=0.0101$, or $p_{1}(u)=h(u;\underline{\mu},\|\mu\|)$ if $\mu$ follows the normal distribution \eqref{eq:mu:normal}, with $\underline{\mu}=4.0907$ and $\|\mu\|=0.2302$ (see table~\ref{table2}).

The probability distribution of inflation instability occurring is
\begin{equation}\label{eq:tube:P2}
P_{2}(\mu_{1})=1-P_{1}(\mu_{1}).
\end{equation}

The probability distributions given by \eqref{eq:tube:P1}-\eqref{eq:tube:P2} are shown in Figure~\ref{fig:tube3} (blue lines for $P_{1}$ and red lines for $P_{2}$). For the deterministic elastic tube, the critical value $\mu_{1}=0.8035\cdot\mu=3.2869$ strictly divides the cases of inflation instability occurring or not. However, in the stochastic case, to increase the chance that inflation is always stable ($P_{1}\approx 1$), one must take sufficiently small values of the parameter $\mu_{1}$, below the expected critical point, while instability is guaranteed ($P_{2}\approx 1$) only for the stochastic neo-Hookean tube.

As an example, in Figure~\ref{fig:tube4}, we show the probability distribution of the normalised internal pressure $T(\lambda)$, defined by \eqref{eq:tube:T}, as a function of the inner stretch $\lambda$, when $\mu$ follows a Gamma distribution with $\rho_{1}=405.0214$ and $\rho_{2}=0.0101$, and $R_{1}=\mu_{1}/\mu$ follows a Beta distribution with $\xi_{1}=287.2297$ and $\xi_{2}=36.1194$ (see table~\ref{table3}). Hence, $\underline{\mu}_{1}=3.6338=0.8883\cdot\underline{\mu}>0.8035\cdot\underline{\mu}$, and instability is expected to occur. However, the probability distribution suggests there is also around 5\% chance that the inflation is stable.

\section{Conclusion}\label{sec:conclude}

For hyperelastic spherical shells and cylindrical tubes under symmetric finite inflation, we showed that, when material parameters are random variables, there is always competition between monotonic expansion and limit-point instability. Specifically, by contrast to the deterministic elastic problem, where there is a single critical value that strictly separate the cases where either the radially symmetric inflation is stable or a limit-point instability occurs, for the stochastic problem, there are probabilistic intervals for the model parameters, where there is a quantifiable chance for both the stable and unstable states to be found.

For numerical illustration, we considered experimental data for rubberlike material, and derived the probability distribution of the corresponding random shear modulus to predict the inflation behaviour of internally pressurised hollow cylinders and spheres made of a material characterised by this parameter. The general framework provided by our stochastic elastic setting is applicable to a class of stochastic hyperelastic materials for which similar results can be obtained. 

This study addresses the need for a deeper understanding of the influence and sources of randomness in natural and industrial materials where mathematical models that take into account the variability in the observed mechanical responses are crucial.


\paragraph{Data access.} The datasets supporting this article have been included in the main text.

\paragraph{Authors contribution.} All authors contributed equally to all aspects of this article and gave final approval for publication.

\paragraph{Competing interests.} The authors declare that they have no competing interests.

\paragraph{Funding.} The support for Alain Goriely by the Engineering and Physical Sciences Research Council of Great Britain under research grant EP/R020205/1 is gratefully acknowledged.


\end{document}